# PRACTICAL APPROACH TO DISTRIBUTED SYSTEMS' DESIGN[1]

Jerzy Mieścicki, Wiktor B. Daszczuk
Waldemar Grabski, Artur Krystosik

Warsaw University of Technology,
Institute of Computer Science

`{jms,wbd,wgr,akr}@ii.pw.edu.pl`

**Summary.** The paper, based on authors' experience from several distributed systems integration projects, summarizes briefly practical designer's view on methodological requirements and overall system organization, including clues as to the organization of the application layer, use of operating system and preferred communication protocols.

## 1. Introduction

The organization of real-time distributed systems, their design methodology and related topics are issues a vast literature is devoted to [e.g. 1, 2, 3]. On the other hand, whenever a practical implementation is considered, the designers usually face the limitations and requirements that are specific to a particular environment (or user) context and that not necessarily conform the computer science state-of-the-art known from textbooks or standards.

The authors of this paper during last two years were (and still are) involved in several research and development projects, done under the auspices of the Institute of Computer Science (Warsaw University of Technology), in a close cooperation with other partners from the WTU (e.g. Institute of Heat Engineering) as well as with non-university partners or sponsors (e.g. Power and Heat Plant in Siekierki, Warsaw, Połaniec Power Plant, German T-Mobil Deutsche Telekom MobilNet GmbH, Mazovian Petrochemical Works in Płock). However different these projects and systems are - they all address basically the same issue: the integration of distributed process control and/or data acquisition systems with new, sophisticated applications designed to run in a new hardware / software environment.

Two of these projects have been just fully implemented and they now these systems work 24 hours per day in a real industrial environment. One of them, consisting in the integration of three different real-time DCSes (Siemens Teleperm, ABB's VMS and UNIX-based system by Hartman and Braun) with the plant management network was implemented for the Siekierki Power and Heat Plant and it was even prized in '96 with a Siemens Award. The other one, designed for the Po³aniec Power Plant, follows the same general design philosophy but with a number of measured variables an order of magnitude greater and with a Proway and

---

[1] This paper was supported by Warsaw University of Technology as a part of PATIA Research Project.

Westinghouse's WDPF fieldbus instead of the three above-mentioned ones. Still other projects (now under development) involve the integration of the ABB's MOD 300 DCS with an original, dedicated measurement and control software which runs in a PC environment and the use of GSM mobile communication protocols for the purposes of widely distributed data acquisition and monitoring systems.

Understandably enough, our research and design team was made to develop an own, practical view on both system organization and the design process itself, based on our experiences on the design and implementation of above-mentioned projects. The goal was to provide a common conceptual framework for a whole class of systems under design and to determine guidelines that could be useful also in the future.

In the present paper, we attempt to summarize our conclusions. In particular, below we attempt:

- to identify the typical methodological requirements applicable to the systems' design,
- to determine the common, overall system structure,
- to determine a general layout of the application layer itself,
- to provide clues as to the use of operating systems, communication protocols as well as software development tools and methodologies,
- to identify the most interesting relevant research problems.

The consecutive chapters of the present paper just follow the above-listed directory of topics.

## 2. Methodological requirements

*Integration of existing system resources and new ones.* Typically, the real-time distributed system designers do not design their system entirely anew, just from the scratch. Most often, there exists a bulk of hardware and software resources that actually work and that have to be just taken for granted. What the user wants is to enhance (or upgrade) the functionality of his system but the requirement is that the existing system components should not be affected and all their operating procedures have to be fully respected.

*Openness to new applications.* Often, the existing distributed control or data acquisition system of a specific plant is delivered by one of renowned, world-scale companies as a closed, turn-key product. Such a large company can, of course, introduce any enhancements or new technological ideas, but at the additional, usually very high cost which may be hardly acceptable. On the other hand, there often exist many local technological or computer companies, university's or institute's research teams etc. which develop still new control algorithms, optimization methods, measurement or signal processing techniques etc. Thus, one important value of our typical project was that it can provide an additional environment for creating new applications which can substantially enhance the system at the low (or even relatively very low) cost.

*Portability and scalability of a project.* From the designers' point of view, the general system organization, communication mechanisms etc., should make the system portable and scalable as much as possible. If this portability and scalability is achieved - each new project could make use of software components from previous ones which could reduce the implementation effort, time and cost.

***General independence of functional features from a physical system architecture.*** One way to achieve design portability and scalability is to design firstly the organization of an application layer (functional software components and their mutual application-layer communication protocols) as if it was generally independent of the organization of a system hardware (with its servers, workstations, terminals, communication links etc.). The mapping of a system software into its hardware structure is then a second step of the implementation. This step allows the designers to determine which particular application-layer protocols have to be implemented as network protocols.

***Use of standard communications protocols*** is an obvious requirement: they are the most broadly used, known, well verified, reliable and guarantee the (relative) independence from physical communications media.

***Use of operating systems and software tools that easily available, affordable and known to the user*** is also an important practical requirement. For instance, it may happen that the designer is tempted to use a real-time operating system (e.g. QNX or OS9) which seems to make an ideal software environment just from designer's point of view, providing real-time multitasking, the explicit influence on resource management mechanisms etc. However, this particular environment and its related libraries, tools etc. may be (and very often it is) rather alien to the user, who may prefer e.g. Windows NT just because it is well known and familiar to him. Also, as a rule, the advanced (and relatively rarely used) systems and tools are much more expensive than the 'popular' ones. One should also take into the consideration the system maintenance issues: as the user is usually allowed to introduce some changes or modifications to the working system - the use of known, familiar software environment may be his important requirement.

## 3. 'Three-dimensional' system model

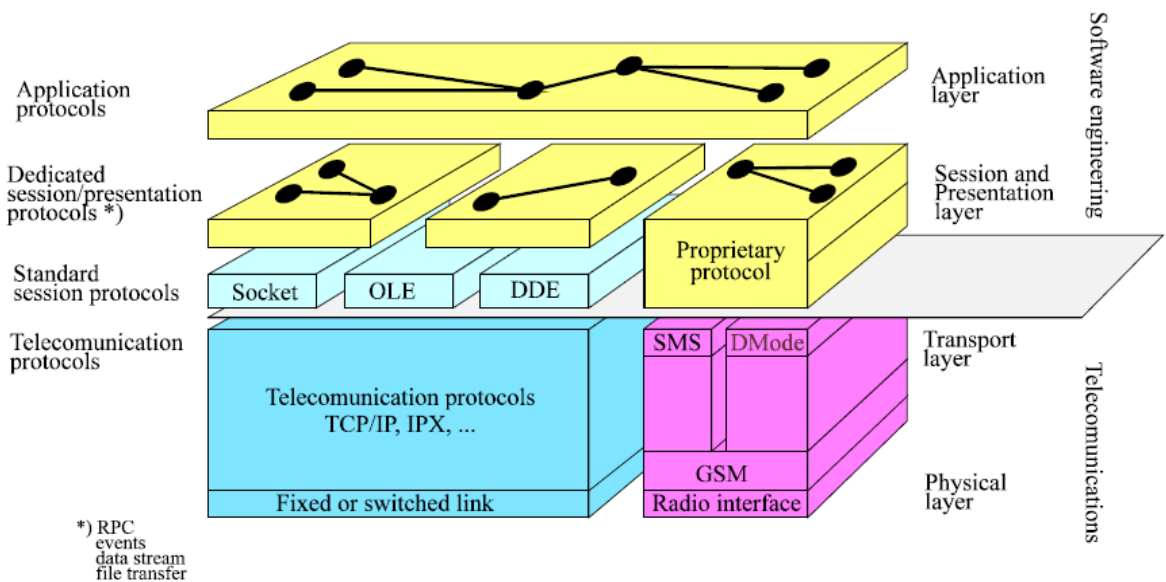

**Fig. 1.** General system organization

### 3.1. Overall system organization

Our approach to a general system organization is illustrated in Fig. 1. The system is viewed as a hierarchy of layers, conforming generally the conventions of ISO OSI Reference Model [2], with the physical link layer as a bottom and the application layer as a top of this hierarchy. Informally, we can say that the transport layer and the underlying region is a typical domain of *telecommunications* while what is above the transport layer - belongs to a typical *software engineering*. The application layer itself has also a specific, typical organization of its own which adds up a 'third dimension' (a 'depth') to the diagram shown in Fig. 1

The main task of a designer is to specify first the application layer software components and to determine virtual protocols among them. Then, the session (and presentation) protocols have to be designed. These in turn can be divided into two sub-layers: The typical elements of the lower of two mentioned sub-layers are *standard session protocols*: e.g. Sockets, OLE or DDE. The *dedicated session / presentation protocols* (which make the upper sub-layer) provide the interface between the application and standard session protocols. They support the typical communication between application components, namely:

- remote procedure call (RPC),
- transfer of events,
- data streams,
- file transfer.

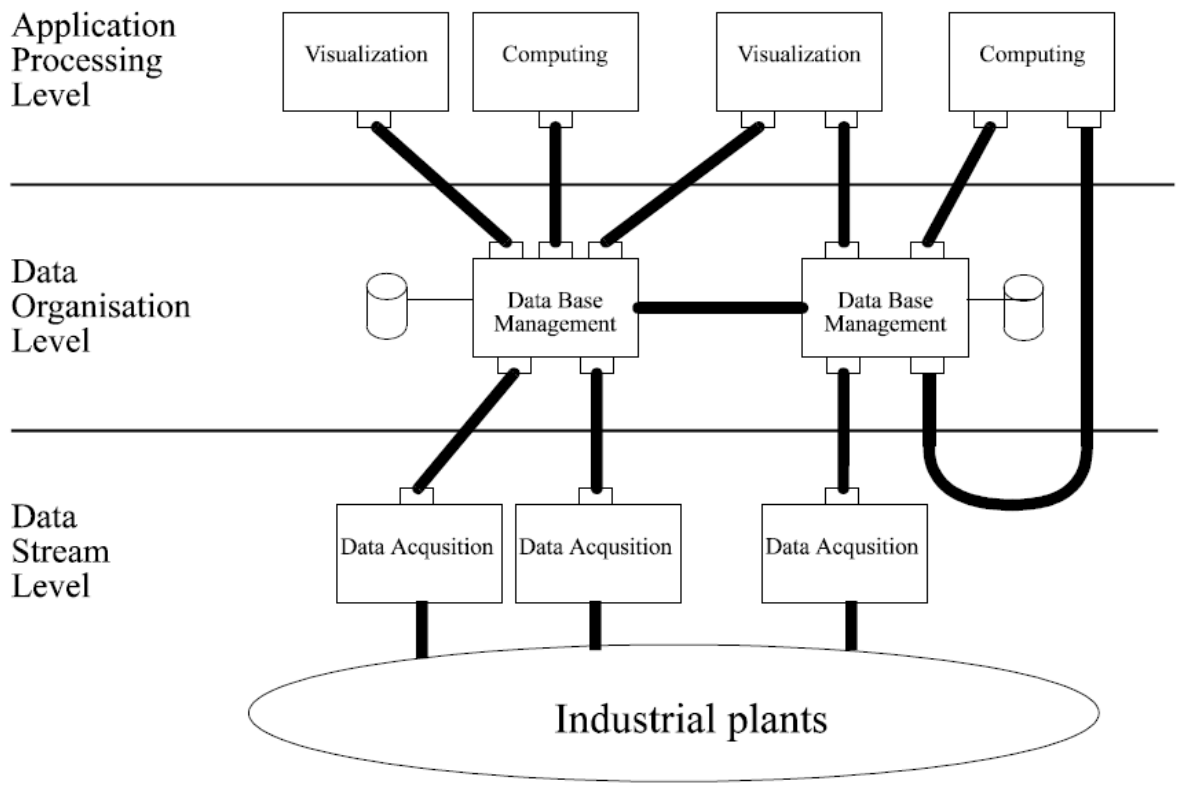

**Fig. 2.** Organization of the application layer

Of course, the use of standard and dedicated session protocols, however recommended, should not be a must. Therefore, the use of own, proprietary session protocols is also allowed.

As it was mentioned in Chapter 2, the mapping of the above functional or logical system organization (covering the application and presentation / session layers which belong to a 'pure' software engineering issues) into the hardware structure of a system allows the designer to determine which session protocols can be implemented immediately (e.g. just via the sockets) and which have to be implemented with the use of telecommunications protocols (e.g. TCP/IP, IPX or other ones, specific to the type of telecommunication subsystem used).

### 3.2. The organization of an application layer

The application layer can be subdivided into three basic *levels* (Fig. 2):

- data stream level which provides interface with the plant DCSes and supports streams of data from (and eventually to) the plant,
- data organization level which supports the storing of data in databases, computation of trends etc.,
- application processing level, consisting of a variety of computing and visualization modules tailored to user's needs (e.g. current plant status displays, reports and statistics on raw materials' usage, engineering calculations supporting operator's or management decisions etc.),

Generally, these three levels can be viewed as an ordered hierarchy, in a sense that the topmost one communicates basically only with the underlying (middle) level but not with the lowest one etc. However, some computing modules can produce their results (or streams of results) and place them in a database exactly the same way as if they were originated at 'raw data' acquisition terminals. Such 'secondary' (or computed) data can be used by other application modules equivalently with 'original' (actually measured) ones. This convention has proved to be very useful in practice.

## 4. Some design decisions

*System platform*. At the early stage of the design the following three types of system platform have been considered as a base for the implementation: Windows (95 and NT), Unix-like systems (including QNX) or Novell NetWare (NLM). Two latter ones (QNX and NLM) are more favorable from designer's point of view: they are 'professional', support real-time multitasking as well as provide the designer the immediate influence on resource management. However, the final decision was to base the design on the Windows NT platform, which is a common and stable one and much better known to users. This facilitates the development of a system itself, makes new future applications easier and solves many maintenance problems.

*Protocols used*. In two projects that have been practically implemented by now, Windows sockets are broadly used as the primary standard session protocols among software components. Also DDE and OLE are used at some extent.

The main network communication protocol was TCP/IP: a known and commonly used one. In one of new projects (now under development) the proprietary protocol based on Short Message Services in a GSM network is developed to provide the system an access to and from mobile data terminals through the radio interface.

*Database implementation issues*. In applications considered by now the main requirement for the database (in a data organization level) was that it has to store the full, real-time stream of most recent data from the plant for a specified time interval (e.g. limited to 10 min.). For longer times, only the trends have to be computed (for specified intervals, e.g. ) and stored in a database for future use. To satisfy this requirement, the real-time database was implemented in a main memory (RAM), as a relatively simple but fast-access data structure. It cooperates with a disk, where the trends from the past are consecutively stored.

## 5. Conclusions

The approach to the system organization presented above has proved to be a convenient framework ordering our 'way of thinking' about both the systems themselves and the practice of their design and implementation. As it was mentioned, two systems following this design philosophy have been implemented yet and still other ones are now under development. Especially, the experience from a *second* implementation has proved that the system organization is pretty portable and scalable. Also, upgrading the system by adding still new engineering computation modules, graphic diagrams, interfaces to other fieldbus types etc. was a relatively easy task. All these advantages have been achieved due to the modularity, the hierarchical approach to communication protocols, assumed independence of functional and physical system structure and the use of standard, most common protocols and software tools.

However, there is also the other side of the coin. Unfortunately, the same factors which cause the above advantages, make the *system's performance* rather obscure. The modularity and the use of standard session protocols as well as network communication protocols always introduces additional communication delays. Moreover, the most commonly used (and therefore convenient and inexpensive) Windows 95 or Windows NT platforms generally do not allow the designer to decide directly upon process scheduling and memory (as well as other resources) management. Thus, the important performance measures (e.g. system response time) can be determined *a posteriori* (i.e. by the performance measurements done in a 'living' system) but they are hardly predictable at the design stage.

In the case of systems designed by now, the time constraints for new applications themselves had not been a serious limitation. All the functions that actually required a 'true' real-time behavior have been implemented within the existing industrial-type DCSes delivered by Siemens, ABB, Westinghouse or Hartmann and Braun. However, in a more general case the system performance evaluation is among the most important elements of a design process.

Therefore, in addition to gaining further experience and know-how on implementation of new systems, the research on performance measurement techniques and systems performance modeling is planned for the near future.